\documentclass[11pt]{article}
\usepackage{indentfirst}
\usepackage{epsfig}
\usepackage{amssymb,amsmath,amsfonts,mathrsfs}
\usepackage{subfigure}
\usepackage{cite}
\usepackage{float}
\usepackage{extarrows}

\newenvironment{mysummary}[1]{%
    \leftskip=2.5em \rightskip=2.5em
    \noindent\small{\bfseries #1}}
    {\par\medskip}
\renewenvironment{abstract}{\begin{mysummary}{Abstract:}}{\end{mysummary}}
\newenvironment{keywords}{\begin{mysummary}{Key words:}}{\end{mysummary}\medskip}
\newenvironment{proof}{\par{\itshape Proof}.\ }
    {\hfill\raisebox{.56ex}{\fbox{}}\par}

\newtheorem{theorem}{\indent Theorem}

\newtheorem{myremark}{\indent Remark}
\newenvironment{remark}{\begin{myremark}\normalfont}
    {\end{myremark}}
\newtheorem{myexample}{\indent Example}

\title{Drag-Tracking Guidance for Entry Vehicles Without Drag Rate Measurement\thanks{This work was supported by the National Nature Science Foundation of China under Grant 61403030.}}
\author{Han Yan\footnote{Engineer, e-mail: yhbice@gmail.com.} and Yingzi He\footnote{Professor.}\\[1ex]
Science and Technology on Space
Intelligent Control Laboratory,\\
Beijing Institute of Control Engineering,
Beijing 100190, China}
\date{}

\begin{document}

\maketitle

\begin{abstract}
A robust entry guidance law without drag rate measurement is designed for drag-tracking in this paper. The bank angle is regarded as the control variable. First, a state feedback guidance law (bank angle magnitude) that requires the drag and its rate as feedback information is designed to make the drag-tracking error be input-to-state stable (ISS) with respect to uncertainties. Then a high gain observer is utilized to estimate the drag rate which is difficult for a vehicle to measure accurately in practice. Stability analysis as well as simulation results show the efficiency of the presented approach.
\end{abstract}


\section{Introduction}
For entry vehicles, guidance algorithm plays an important role in steering the vehicle through the atmospheres safely with mission requirements. In general, guidance methods for entry can be classified into two types: predictor-corrector guidance and reference-trajectory guidance. The main merit of predictor-corrector guidance is that it can update the reference trajectory every guidance period online so as to improve the guidance precision. But computing reference trajectory online is time consuming, so the method may be not feasible in practice.  In particular, the performance of predictor-corrector guidance might be degraded if we do not have a good grasp of the atmosphere information (such as the Mars atmosphere), since the scheme relies on the accurate entry dynamics model \cite{Guo_2013}. The drag acceleration is strongly related to the measurable accelerations, and it further has exact kinematic relationship with the arc length of the flying trajectory \cite{Saraf_2004}. Therefore, the drag profile tracking approach, which is one kind of reference-trajectory guidance, has been validated in the Apollo and Shuttle Programs \cite{Shuttle} and extensively investigated, and comparing with the predictor-corrector guidance, the approach has distinct advantages in realization.

The feedback linearization method is a typical tool that was applied in the drag-tracking control \cite{Leavitt_2007,Mease_2002,Wang_2013,Saraf_2004,Talole_2007}, but the desired asymptotic convergence of the drag tracking error can not be guaranteed in the presence of uncertainties or control saturation. In order to take the control saturation into account, Lu and co-workers \cite {Lu_1994} solved the tracking control problem for a continuous-time nonlinear system with bounded input by using the continuous-time nonlinear predictive control method, and they applied the approach to drag-tracking problem \cite{Lu_1997}. In their works, the control was designed to minimize a cost function related to the predicted error. \cite{Benito_2008} used a different cost function to design the control law for drag-tracking. Considering the model uncertainties, the guidance law design problem for low-lifting skip reentry subject to control saturation was studied in \cite{Guo} based on nonlinear predictive control in the case that the number of output variables does not equal the number of input variables. The robust control that was applied to robot manipulators \cite{Spong} was also adopted for drag-tracking in Mars atmospheric entry flight to make the tracking error converge into a small neighbourhood of zero \cite{Guo_2013}. Most of drag-tracking guidance laws (e.g. see \cite{Guo_2013,Saraf_2004,Lu_1997,Benito_2008,Guo}) require the knowledge of drag rate, which is hard for a vehicle to measure accurately in practice. Thus, the altitude rate was used as feedback instead of the drag rate in the Shuttle guidance, but it is error prone as mentioned in \cite{Shuttle}. The sliding mode state and perturbation observer (SMSPO) was used in \cite{Talole_2007} to address the issue of estimation of the drag rate. In \cite{Xia_2013}, the active disturbance rejection control (ADRC) algorithm was utilized to design a drag-tracking law for Mars pinpoint landing, and an extended state observer (ESO) was introduced to estimate the drag rate and a extended state. However, uncertainties were not fully considered in \cite{Talole_2007,Xia_2013}, and the stability analysis of the closed-loop system in \cite{Xia_2013} was not provided.
Besides the idea of tracking reference drag, \cite{Tian_2011} studied the full states trajectory tracking control problem under the multi-constrained conditions by using Legendre pseudospectral feedback method.

In this paper, a robust entry guidance law without drag rate measurement is designed for drag-tracking. First, the drag dynamics with uncertainties is formulated, and a state feedback guidance law (desired bank angle magnitude), which requires the drag and its rate as feedback information, is designed to make the drag-tracking error be input-to-state stable (ISS) with respect to uncertainties. As stated earlier, the drag rate is not feasible to be used as feedback information, and in view of this, a high-gain observer \cite{Khalil_1992,Khalil_1999} is integrated into the guidance law to estimate the drag rate, which fulfills the guidance law design without drag rate measurement. Moreover, the stability analysis is provided to show that ISS property of the closed-loop system under the state feedback guidance law can be recovered by using a sufficiently fast high-gain observer.

The remainder of this paper is organized as follows. The drag dynamics is formulated in Section \ref{sec of model}. After presenting the state feedback guidance law, the high-gain observer is introduced to estimate the drag rate in Section \ref{guidance law design}. Section \ref{simulation} shows the simulation results. Finally, Section \ref{conclusion} summarizes the conclusions.

\section{Model Derivation}\label{sec of model}
The motion equations of an unpowered, point mass vehicle flying over a non-rotating planet in a stationary atmosphere are given by \cite{Talole_2007,Xia_2013,Tian_2011,Guo}
\begin{subequations}
\begin{equation}\label{dot_r}
\dot{r}=v\sin\gamma
\end{equation}
\begin{equation}
\dot{\phi}=\frac{v\cos\gamma\sin\chi}{r\cos\theta}
\end{equation}
\begin{equation}
\dot{\theta}=\frac{v\cos\gamma\cos\chi}{r}
\end{equation}
\begin{equation}\label{dot_v}
\dot{v}=-D-g\sin\gamma
\end{equation}
\begin{alignat}{1}
\dot{\gamma}=&\frac{L\cos\sigma}{v}-\left(\frac{g}{v}-\frac{v}{r}\right)\cos\gamma
\end{alignat}
\begin{equation}
\dot{\chi}=\frac{L\sin\sigma}{v\cos\gamma}+\frac{v\cos\gamma\sin\chi\tan\theta}{r}
\end{equation}
\end{subequations}
where $r$ is the radial position, $\phi$ is longitude, $\theta$ latitude, $v$ is the velocity, $\gamma$ is the flight path angle, $\chi$ is the heading angle, $L$ is the lift acceleration, $D$ is the drag acceleration, and $g$ is gravitational acceleration. $L$ and $D$ can be calculated as
\begin{subequations}
\begin{equation}
L=\frac{1}{2m}\rho v^{2}S(\underbrace{C_{L}^{0}+\Delta C_{L}}_{C_{L}})
\end{equation}
\begin{equation}\label{eq of drag}
D=\frac{1}{2m}\rho v^{2}S(\underbrace{C_{D}^{0}+\Delta C_{D}}_{C_{D}})
\end{equation}
\end{subequations}
where $m$ is the vehicle mass, $\rho$ is the atmospheric density, $S$ is the reference area, $C_{L}^{0}$ and $C_{D}^{0}$ are nominal values of aerodynamic coefficients, and $\Delta C_{L}$ and $\Delta C_{D}$ are bounded uncertainties. An exponential atmospheric density model
\begin{equation}\label{exponential atmospheric density model}
\rho=\rho_{0}e^{-\frac{h}{h_{s}}}+\Delta \rho
\end{equation}
is assumed, where $h=r-r_{0}$, $r_{0}$ is the reference radius, $\rho_{0}$ is atmospheric density at the reference radius, $\Delta \rho$ is bounded uncertainty, and $h_{s}$ is characteristic constant. The gravitational acceleration as a function of $r$ is given by
\begin{equation}
g=\frac{\mu}{r^{2}}
\end{equation}
where $\mu$ is gravitational constant.

\section{Guidance Law Design}\label{guidance law design}
\subsection{State Feedback Guidance Law Based on ISS}

Due (\ref{eq of drag}), one has
\begin{equation}\label{dot_D}
\dot{D}=\frac{1}{2}\dot{\rho}v^{2}C_{D}\frac{S}{m}+\rho v\dot{v}C_{D}\frac{S}{m}+\frac{1}{2}\rho v^{2}\dot{C}_{D}\frac{S}{m}
\end{equation}
and
\begin{equation}
\frac{\dot{D}}{D}=\frac{\dot{\rho}}{\rho}+\frac{2\dot{v}}{v}+\frac{\dot{C}_{D}}{C_{D}}
\end{equation}
It can also be calculated out that
\begin{subequations}
\begin{equation}\label{dot_rho}
\frac{\dot{\rho}}{\rho}=-\frac{\dot{h}}{h_{s}}+\delta_{\rho}=-\frac{\dot{r}}{h_{s}}+\delta_{\rho}\xlongequal{\mathrm{Eq.}~(\ref{dot_r})}-\frac{v\sin\gamma}{h_{s}}+\delta_{\rho}
\end{equation}
\begin{equation}\label{dot_C_D}
\frac{\dot{C}_{D}}{C_D}=\frac{\dot{C}_{D}^{0}}{C_D^{0}}+\delta_{C_D}
\end{equation}
\begin{equation}
\dot{g}=-\frac{2\mu}{r^{3}}v\sin\gamma=-\frac{2gv\sin\gamma}{r}
\end{equation}
\end{subequations}
where $\delta_{\rho}=\frac{\rho\Delta\dot{\rho}-\dot{\rho}\Delta\rho}{\rho(\rho-\Delta\rho)}$ and $\delta_{C_D}=\frac{\Delta\dot{C}_{D}C_{D}^{0}-\Delta C_{D}\dot{C}_{D}^{0}}{C_{D}^{0}(C_{D}^{0}+\Delta C_{D})}$. Thus,
\begin{equation}\label{dotD}
\frac{\dot{D}}{D}=-\frac{v\sin\gamma}{h_s}-\frac{2D}{v}-\frac{2g\sin\gamma}{v}+\underbrace{\frac{\dot{C}_{D}^{0}}{C_{D}^{0}}}_{C}+\underbrace{\delta_{\rho}+\delta_{C_D}}_{\delta}
\end{equation}
Furthermore,
\begin{equation}\label{ddotD}
\ddot{D}=f(D,t)+g_{0}(D,t)u+\Delta(D,t)
\end{equation}
where
\begin{equation}
u=\cos\sigma
\end{equation}
and
\begin{subequations}\nonumber
\footnotesize{\begin{alignat}{1}
f=&\left(-\frac{v\sin\gamma}{h_{s}}-\frac{4D}{v}-\frac{2g\sin\gamma}{v}+C\right)\left(-\frac{v\sin\gamma}{h_s}D-\frac{2D^2}{v}-\frac{2g\sin\gamma}{v}D+CD\right)\nonumber\\
&+D\left(\frac{D\sin\gamma+g}{h_{s}}+\frac{4g\sin^{2}\gamma-2g\cos^{2}\gamma}{r}-\frac{2D^{2}+4Dg\sin\gamma+2g^{2}\sin^{2}\gamma-2g^{2}\cos^{2}\gamma}{v^2}\right.\left.+\frac{v^{2}\cos^{2}\gamma}{rh_{s}}+\dot{C}\right)\nonumber
\end{alignat}}
\begin{equation}\nonumber
g_{0}=-\left(\frac{v}{h_{s}}+\frac{2g}{v}\right)\frac{LD\cos\gamma}{v}
\end{equation}
\begin{equation}\nonumber
\Delta=\dot{\delta}D+\left(-\frac{2v\sin\gamma}{h_{s}}-\frac{2D^{2}}{v}-\frac{4D}{v}-\frac{4g\sin\gamma}{v}+C+CD+\delta D\right)\delta
\end{equation}
\end{subequations}
Since the purpose of designing a guidance law is to make the drag acceleration $D$ track its reference value $D^{*}$ by modulating the bank angle $\sigma$, we define $\widetilde{D}=D-D^{*}$ and $x=[x_{1},x_{2}]^{T}=[\widetilde{D},\dot{\widetilde{D}}]^{T}$. The drag dynamics for guidance law design is formulated as
\begin{alignat}{1}\label{guidance system}
\dot{x}=\begin{bmatrix}
x_{2}\\
f(D,t)-\ddot{D}^{*}
\end{bmatrix}+\begin{bmatrix}
0\\
g_{0}(D,t)
\end{bmatrix}u+\begin{bmatrix}
0\\
\Delta(D,t)
\end{bmatrix}
\end{alignat}
Here, we assume that uncertainties $\delta$ and $\dot{\delta}$ are bounded, and in a reasonable flight domain of interest there exist positive constants $l$ and $d$ such that
\begin{equation}\label{assumption}
|\Delta|\leq l|x_{1}|+d
\end{equation}
holds. In practice, the flight path angle $\gamma$ always satisfies $-90^{\circ}<\gamma<90^{\circ}$. From this, clearly, $g_0$ is invertible.
Regarding (\ref{guidance system}), we have the following theorem.
\begin{theorem}\label{theroem for state feedback}
Consider the system (\ref{guidance system}). There exists a guidance law
\begin{equation}\label{guidance law}
u=g_{0}^{-1}\left(-f+\ddot{D}^{*}-\frac{a}{\varepsilon^{2}_{0}}x_{1}-\frac{b}{\varepsilon_{0}}x_{2}\right)
\end{equation}
with $a>0$, $b>0$ and $\varepsilon_{0}>0$, such that the closed-loop system is ISS with respect to $d$, and
moreover, the influence of uncertainties on $x$ can be made close to zero for sufficiently small $\varepsilon_{0}$.
\end{theorem}
\begin{proof}
Substituting guidance law (\ref{guidance law}) into (\ref{guidance system}) yields
\begin{equation}\label{guidance system with guidance law}
\dot{x}=\underbrace{\begin{bmatrix}
0 & 1\\
-\frac{a}{\varepsilon^{2}_{0}} & -\frac{b}{\varepsilon_{0}}
\end{bmatrix}}_{F}x+\underbrace{\begin{bmatrix}
0\\
1
\end{bmatrix}}_{B}\Delta
\end{equation}
The change of variables
\begin{equation}\label{transformation in theorem 1}
\zeta_{1}=\frac{x_{1}}{\varepsilon_{0}},~\zeta_{2}=x_{2}
\end{equation}
brings (\ref{guidance system with guidance law}) into the form
\begin{equation}\label{guidance system with guidance law in theorem 1}
\varepsilon_{0}\dot{\zeta}=\underbrace{\begin{bmatrix} 0 & 1\\
-a & -b
\end{bmatrix}}_{F_{0}}\zeta+\varepsilon_{0} B\Delta
\end{equation}
where $\zeta=[\zeta_{1},\zeta_{2}]^{T}$ and
$F_{0}$ is a Hurwitz matrix. The derivative of Lyapunov function
\begin{equation}\label{V for guidance}
V(\zeta)=\zeta^{T}P_{0}\zeta
\end{equation}
where $P_{0}$ is the positive definite solution of the Lyapunov equation $P_{0}F_{0}+F_{0}^{T}P_{0}=-I$, along the trajectories of system (\ref{guidance system with guidance law in theorem 1}) is given by
\begin{alignat}{1}\label{dot(V)}
\dot{V}&=-\frac{1}{\varepsilon_{0}}\|\zeta\|^{2}+2\zeta^{T}P_{0}B\Delta\nonumber\\
&\leq -\frac{1}{\varepsilon_{0}}\|\zeta\|^{2}+2\|\zeta\|\|P_{0}B\|(\varepsilon_{0}l|\zeta_{1}|+d)
\end{alignat}
Substituting the inequalities
$$
\zeta^{T}P_{0}Bd\leq \frac{1}{2}\|\zeta\|^{2}+\frac{1}{2}\|P_{0}\|^{2}\|B\|^{2}d^{2}
$$
$$
\varepsilon_{0}\|\zeta\|\|P_{0}B\|l|\zeta_{1}|\leq \varepsilon_{0}\|P_{0}B\|l\|\zeta\|^{2}
$$
into Eq. (\ref{dot(V)}), we obtain
\begin{equation}\label{dot(V) with inequality}
\dot{V}\leq -\underbrace{\left(\frac{1}{\varepsilon_{0}}-1-2\varepsilon_{0}\|P_{0}B\|l\right)}_{\kappa(\varepsilon_{0})}\|\zeta\|^{2}+\|P_{0}\|^{2}\|B\|^{2}d^{2}
\end{equation}
The boundedness of $l$ leads to the fact that $\kappa(\varepsilon_{0})>0$ for sufficiently small $\varepsilon_{0}$, and in this case, since
\begin{equation}\label{inequality of V}
\lambda_{\min}(P_{0})\|\zeta\|^{2}\leq V(\zeta)\leq \lambda_{\max}(P_{0})\|\zeta\|^{2}
\end{equation}
we have
\begin{equation}\label{solution of dot(V)}
\|\zeta(t)\|^{2}\leq \lambda_{3}e^{-\lambda_{1}t}\|\zeta(0)\|^{2}+\frac{\lambda_{2}}{\lambda_{1}}(1-e^{-\lambda_{1}t})d^{2}
\end{equation}
and
\begin{alignat}{1}
\|x(t)\|= \|\varphi(\varepsilon_{0})\zeta(t)\|
\leq\sqrt{\lambda_{3}e^{-\lambda_{1}t}}\|x(0)\|+\|\varphi(\varepsilon_{0})\|\sqrt{\frac{\lambda_{2}}{\lambda_{1}}}d\label{ineq of x in ISS form}
\end{alignat}
where $\varphi(\varepsilon_{0})=\mathrm{diag}(\varepsilon_{0},1)$, $\lambda_{1}=\frac{\kappa(\varepsilon_{0})}{\lambda_{\max}(P_{0})}$, $\lambda_{2}=\frac{\|B\|^{2}\|P_{0}\|^{2}}{\lambda_{\min}(P_{0})}$, $\lambda_{3}=\frac{\lambda_{\max}(P_{0})}{\lambda_{\min}(P_{0})}$.
From Eq. (\ref{ineq of x in ISS form}), it can be seen that the closed-loop system is ISS with respect to $d$, and the influence of uncertainties on $x$ will be close to zero for sufficiently small $\varepsilon_{0}$.

\end{proof}

If all the variables can be measured accurately, $\dot{D}$ can be calculated from Eq. (\ref{dotD}). However, there are always unknown uncertainties in atmospheric density and aerodynamic coefficients \cite{Guo,Guo_2013}, i.e., $\delta\neq0$, which implies that the accurate information of $\dot{D}$ is hard to get actually. The guidance law without drag rate measurement will be designed in next subsection by combining a high-gain observer with guidance law (\ref{guidance law}).

\begin{remark}
A similar result (ISS property of the close-loop system) has been got in \cite{Guo} under the assumption that the uncertainties related term $\Delta$ is bounded. However, since $\Delta$ is also a function of $\widetilde{D}$, strictly speaking, the boundedness of $\Delta$ cannot be guaranteed. Different from \cite{Guo}, we assume that $|\Delta|$ is not bigger than a linear function of $|\widetilde{D}|$ as shown by (\ref{assumption}) in a reasonable flight domain of interest, and a robust guidance law is also obtain based on ISS theory.
\end{remark}

\subsection{Guidance Law without Drag Rate}
The guidance law without knowledge of drag rate can be got by replacing $\hat{x}_{2}$ instead of $x_{2}$ in (\ref{guidance law}), i.e.,
\begin{equation}\label{guidance law with observer}
u=g_{0}^{-1}\left(-f+\ddot{D}^{*}-\frac{a}{\varepsilon^{2}_{0}}x_{1}-\frac{b}{\varepsilon_{0}}\hat{x}_{2}\right)
\end{equation}
where $\hat{x}_{2}$ is the estimate of drag rate and generated by the high-gain observer
\begin{subequations}\label{observer}
\begin{equation}
\dot{\hat{x}}_{1}=\hat{x}_{2}+\frac{l_{1}}{\varepsilon}(x_{1}-\hat{x}_{1})
\end{equation}
\begin{equation}
\dot{\hat{x}}_{2}=-\frac{a}{\varepsilon^{2}_{0}}x_{1}-\frac{b}{\varepsilon_{0}}\hat{x}_{2}+\frac{l_{2}}{\varepsilon^{2}}(x_{1}-\hat{x}_{1})
\end{equation}
\end{subequations}
with $l_{1}>0$, $l_{2}>0$, $\varepsilon>0$, and $\hat{x}=[\hat{x}_{1},\hat{x}_{2}]^{T}$.
The main results can be stated as the following theorem.
\begin{theorem}\label{theorem with observer}
Consider the closed-loop system of system (\ref{guidance system}) and guidance law (\ref{guidance law with observer}) with high-gain observer (\ref{observer}). Let $\tilde{x}=x-\hat{x}$.
There exists a positive constant $\varepsilon_{1}^{*}$ such that, for every $0<\varepsilon<\varepsilon_{1}^{*}$, $(x,\tilde{x})$ is ISS with respect to $d$, and the uncertainties can be suppressed by adjusting $\varepsilon$ and $\varepsilon_{0}$. Besides, if $d$ vanishes, there exists $\varepsilon^{*}_{2}>0$ such that, for every $0<\varepsilon<\varepsilon_{2}^{*}$, $x$ and $\tilde{x}$ can converge to zero exponentially.
\end{theorem}
\begin{proof}
The change of variables
\begin{equation}\label{transform in theorem 2}
\eta_{1}=\frac{\tilde{x}_{1}}{\varepsilon},~\eta_{2}=\tilde{x}_{2}
\end{equation}
bring the closed-loop system into the form
\begin{subequations}
\begin{equation}
\dot{x}=Fx+B\left(\Delta+\frac{b}{\varepsilon_{0}}\eta_{2}\right)
\end{equation}
\begin{equation}
\varepsilon\dot{\eta}=\underbrace{\begin{bmatrix}
-l_{1} & 1\\
-l_{2} &0
\end{bmatrix}}_{A_{0}}\eta+\varepsilon B\Delta
\end{equation}
\end{subequations}
where $\eta=[\eta_{1},\eta_{2}]^{T}$, $F$ and $B$ have been given in (\ref{guidance system with guidance law}), and $A_{0}$ is a Hurwitz matrix.
Transforming $x$ to $\zeta$ by Eq. (\ref{transformation in theorem 1}), we rewritten the closed-loop system as
\begin{subequations}\label{compact singularly perturbed form}
\begin{equation}\label{compact singularly perturbed form1}
\dot{\zeta}=\frac{1}{\varepsilon_{0}}F_{0}\zeta+B\left(\Delta+\frac{b}{\varepsilon_{0}}\eta_{2}\right)
\end{equation}
\begin{equation}\label{compact singularly perturbed form2}
\dot{\eta}=\frac{1}{\varepsilon}A_{0}\eta+B\Delta
\end{equation}
\end{subequations}
The derivative of Lyapunov function
\begin{equation}\label{V for output feedback guidance}
V_{com}(\zeta,\eta)=V(\zeta)+\eta^{T}P\eta
\end{equation}
where $V(\zeta)$ is defined by (\ref{V for guidance}) and $P$ is the positive definite solution of the Lyapunov equation $PA_{0}+A_{0}^{T}P=-I$, along the trajectories of system (\ref{compact singularly perturbed form}) is given by
\begin{alignat}{1}
\dot{V}_{com}=&-\frac{1}{\varepsilon_{0}}\|\zeta\|^{2}+2\zeta^{T}P_{0}B\left(\Delta+\frac{b}{\varepsilon_{0}}\eta_{2}\right)
-\frac{1}{\varepsilon}\|\eta\|^{2}+2\eta^{T}PB\Delta\nonumber\\
\leq & -\frac{1}{\varepsilon_{0}}\|\zeta\|^{2}+2\|\zeta\|\|P_{0}B\|\left(\varepsilon_{0}l|\zeta_{1}|+d+\frac{b}{\varepsilon_{0}}|\eta_{2}|\right)-\frac{1}{\varepsilon}\|\eta\|^{2}+2\|\eta\|\|PB\|(\varepsilon_{0}l|\zeta_{1}|+d)\nonumber\\
\leq &-\frac{1}{\varepsilon_{0}}\|\zeta\|^{2}+2\|\zeta\|\|P_{0}B\|\left(\varepsilon_{0}l\|\zeta\|+d+\frac{b}{\varepsilon_{0}}\|\eta\|\right)-\frac{1}{\varepsilon}\|\eta\|^{2}+2\|\eta\|\|PB\|(\varepsilon_{0}l\|\zeta\|+d)
\end{alignat}
Substituting the inequality
\begin{subequations}\nonumber
\begin{equation}\nonumber
\|\zeta\|\|P_{0}B\|d\leq \frac{1}{2}\|\zeta\|^{2}+\frac{1}{2}\|P_{0}\|^{2}\|B\|^{2}d
\end{equation}
\begin{equation}\nonumber
\|\eta\|\|PB\|d\leq \frac{1}{2}\|\eta\|^{2}+\frac{1}{2}\|P\|^{2}\|B\|^{2}d
\end{equation}
\end{subequations}
yields
\begin{alignat}{1}
\dot{V}_{com}\leq&-\underbrace{\left(\frac{1}{\varepsilon_{0}}-1-2\varepsilon_{0}\|P_{0}B\|l\right)}_{\kappa(\varepsilon_{0})}\|\zeta\|^{2}-\left(\frac{1}{\varepsilon}-1\right)\|\eta\|^{2}\nonumber\\
&+2\underbrace{\left(\frac{b}{\varepsilon_{0}}\|P_{0}B\|+\varepsilon_{0}l\|PB\|\right)}_{\alpha}\|\zeta\|\|\eta\|+\underbrace{(\|P_{0}\|^{2}+\|P\|^{2})\|B\|^{2}}_{C_{0}}d^{2}\nonumber\\
=&-\mathcal{X}^{T}Q\mathcal{X}+C_{0}d^{2}
\end{alignat}
where
\begin{equation}\nonumber
\mathcal{X}=\begin{bmatrix}
\|\zeta\|\\
\|\eta\|
\end{bmatrix},~
Q=\begin{bmatrix}
\kappa(\varepsilon_{0}) & -\alpha\\
-\alpha & \frac{1}{\varepsilon}-1
\end{bmatrix}
\end{equation}
For bounded $l$ and $\kappa(\varepsilon_{0})>0$, the matrix $Q$ will be positive define for sufficiently small $\varepsilon$. Hence, there exists $\varepsilon_{1}^{*}>0$ such that, for $0<\varepsilon<\varepsilon_{1}^{*}$, we have $\lambda_{\min}(Q)>0$, and the inequality
\begin{equation}\label{composite Lyapunov function}
\dot{V}_{com}\leq-\lambda_{\min}(Q)\|\mathcal{X}\|^{2}+C_{0}d^{2}
\end{equation}
holds. Let $\mathcal{Y}=[\zeta,\eta]^{T}$. Since $\|\mathcal{X}\|=\|\mathcal{Y}\|$, we have
\begin{equation}\label{inequality of V_(c)}
\lambda_{\min}(P^{'})\|\mathcal{X}\|^{2}=\lambda_{\min}(P^{'})\|\mathcal{Y}\|^{2}\leq V_{com}=\mathcal{Y}^{T}P^{'}\mathcal{Y}\leq \lambda_{\max}(P^{'})\|\mathcal{Y}\|^{2}=\lambda_{\max}(P^{'})\|\mathcal{X}\|^{2}
\end{equation}
where $P^{'}=\mathrm{block~diag}\{\frac{1}{2}I_{2},P_{0}\}$. Substituting Eq. (\ref{inequality of V_(c)}) into Eq. (\ref{composite Lyapunov function}) yields
\begin{equation}
\dot{V}_{com}\leq -\frac{\lambda_{\min}(Q)}{\lambda_{\max}(P^{'})}V_{com}+C_{0}d^{2}
\end{equation}
that is
\begin{equation}\label{solution of dot(V_(c))}
V_{com}(\mathcal{Y}(t))\leq e^{-\lambda_{1} t}V_{com}(\mathcal{Y}(0))+\frac{C_{0}}{\lambda_{1}}(1-e^{-\lambda_{1} t})d^{2}
\end{equation}
where $\lambda_{1}=\frac{\lambda_{\min}(Q)}{\lambda_{\max}(P^{'})}$. Substituting Eq. (\ref{inequality of V_(c)}) into Eq. (\ref{solution of dot(V_(c))}), we have
\begin{equation}\label{solution of dot(V_(c)) with inequality of V_(c)}
\|\mathcal{Y}(t)\|^{2}\leq \lambda_{2}e^{-\lambda_{1} t}\|\mathcal{Y}(0)\|^{2}+\frac{C_{0}}{\lambda_{1}\lambda_{3}}(1-e^{-\lambda_{1} t})d^{2}
\end{equation}
where $\lambda_{2}=\frac{\lambda_{\max}(P^{'})}{\lambda_{\min}(P^{'})},~\lambda_{3}=\lambda_{\min}(P^{'})$.
Therefore, $\mathcal{Y}(t)$ satisfies
\begin{equation}\label{state of closed-loop system in the similar form of ISS}
\|\mathcal{Y}(t)\|\leq \sqrt{\lambda_{2}e^{-\lambda_{1} t}}\|\mathcal{Y}(0)\|+\sqrt{\frac{C_{0}}{\lambda_{1}\lambda_{3}}(1-e^{-\lambda_{1} t})}d
\end{equation}
Therefore, system (\ref{compact singularly perturbed form}) is ISS with respect to $d$ for $0<\varepsilon\leq \varepsilon^{*}_{1}$. $\lambda_{1}$ can be sufficiently large by adjusting $\varepsilon_{0}$ and $\varepsilon$, and, accordingly, it can be seen from Eq. (\ref{state of closed-loop system in the similar form of ISS}) that the uncertainties can be suppressed.

Moreover, if $d$ vanishes, (\ref{state of closed-loop system in the similar form of ISS}) can be rewritten as $\|\mathcal{Y}(t)\|\leq \sqrt{\lambda_{2}e^{-\lambda_{1} t}}\|\mathcal{Y}(0)\|$, where $\lambda_{1}>0$ for sufficiently small $\varepsilon$. Therefore, there exists $\varepsilon_{2}^{*}>0$ such that, for every $0<\varepsilon\leq \varepsilon^{*}_{2}$, the origin of system (\ref{compact singularly perturbed form}) is exponentially stable.

The proof is completed.
\end{proof}


\begin{remark}
Actually, $u=\cos\sigma$ is bounded by $\pm1$, but the control saturation problem is not considered in this paper since it is implicitly assumed that the vehicle has enough maneuvering capacity to achieve drag-tracking in reasonable cases by modulating the bank angle magnitude ($0^{\circ}\leq\sigma\leq180^{\circ}$). Guidance law design with measurable information for entry subject to control saturation will be investigated in the future work.
\end{remark}

\section{Simulation Results}\label{simulation}
This section presents simulation results to test the performance of the proposed guidance laws.

Consider the Mars atmospheric entry flight, and vehicle, reference drag profile and other data from \cite{Guo_2013} are used. The Mars
lander has surface area of $16\mathrm{m^{2}}$ and weighs 992kg \cite{Mars}. The lift-to-drag ratio and the ballistic coefficient are 0.18 and $115\mathrm{kg/m^{2}}$, respectively. The initial and final state variables can be found in Table \ref{State}. It can be calculated out that the desired total downrange is 723.32km.

\begin{figure}[H]
  \centering
  \subfigure[Drag acceleration]{
    \includegraphics[width=2.5in]{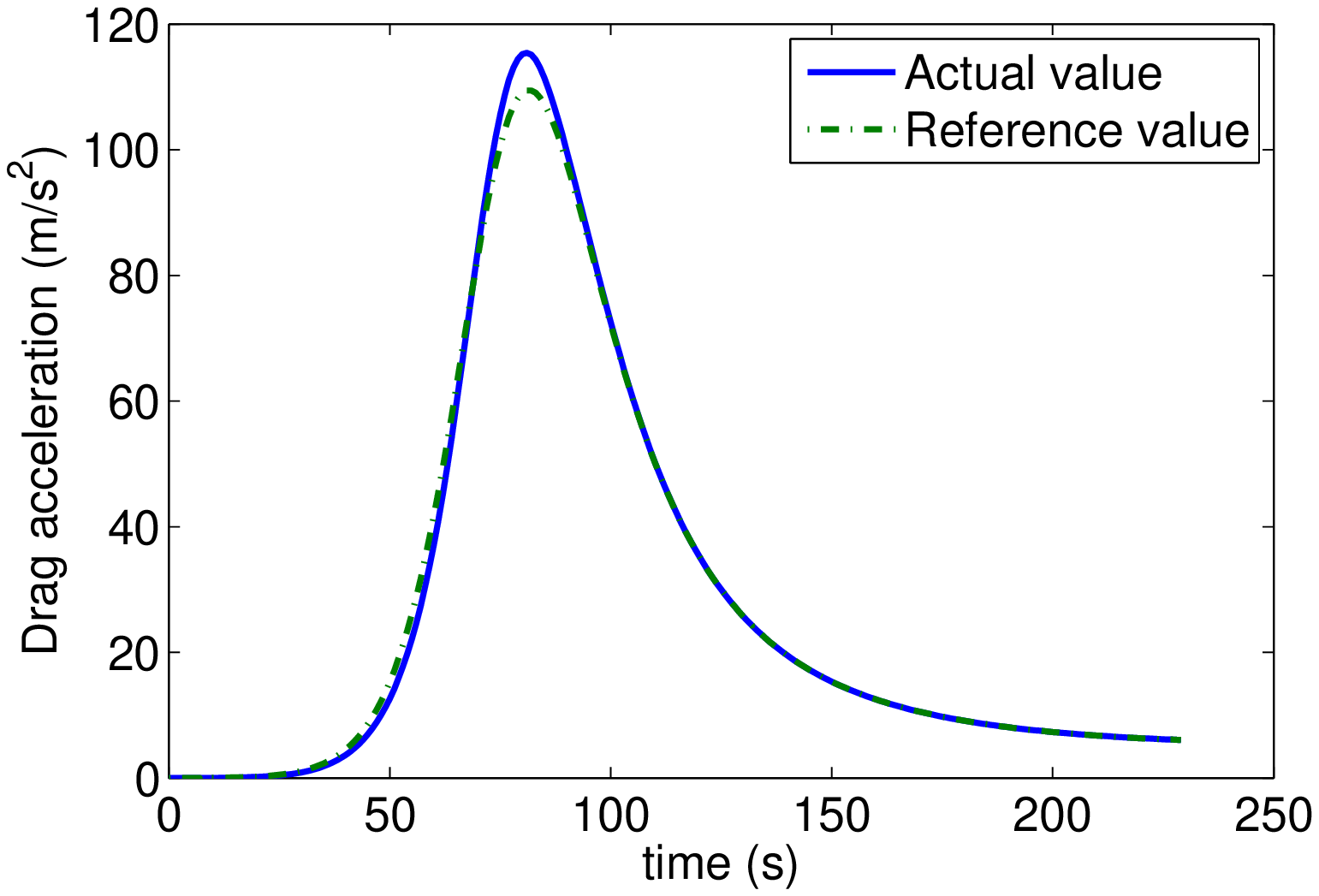}}
  \subfigure[Drag acceleration tracking error]{
    \includegraphics[width=2.5in]{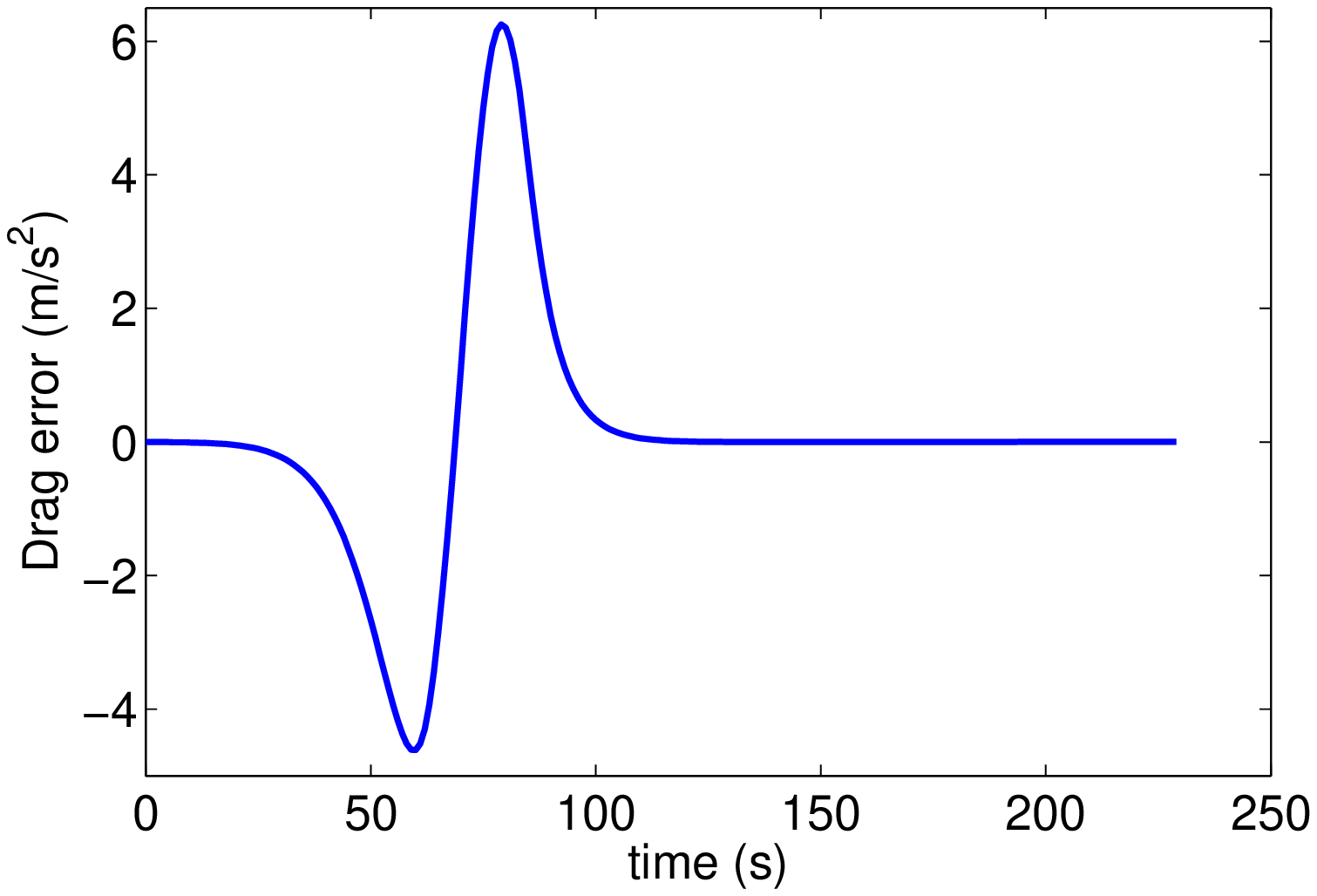}}
  \caption{Drag tracking with state feedback guidance law (\ref{guidance law})\label{drag}}
\end{figure}

\begin{figure}[H]
  \centering
  \subfigure[Altitude]{
    \includegraphics[width=2.5in]{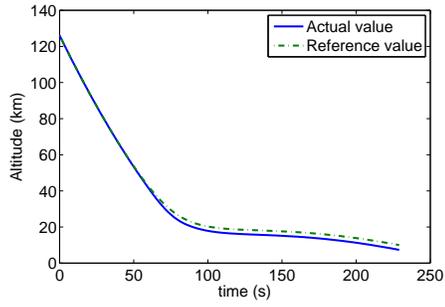}}
  \subfigure[Altitude tracking error]{
    \includegraphics[width=2.5in]{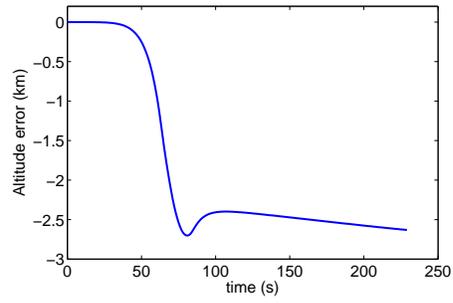}}
  \caption{Altitude tracking with state feedback guidance law (\ref{guidance law})\label{Altitude}}
\end{figure}

\begin{figure}[H]
  \centering
  \subfigure[Velocity]{
    \includegraphics[width=2.5in]{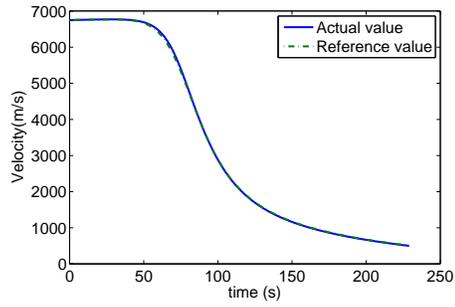}}
  \subfigure[Velocity tracking error]{
    \includegraphics[width=2.5in]{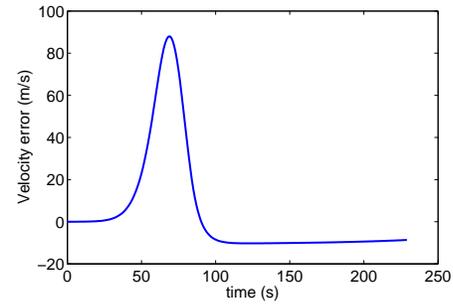}}
  \caption{Velocity tracking with state feedback guidance law (\ref{guidance law})\label{Velocity}}
\end{figure}

\begin{figure}[H]
  \centering
  \subfigure[$\sigma$]{
    \includegraphics[width=2.5in]{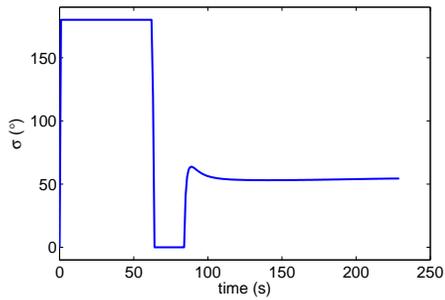}}
  \subfigure[$\gamma$]{
    \includegraphics[width=2.5in]{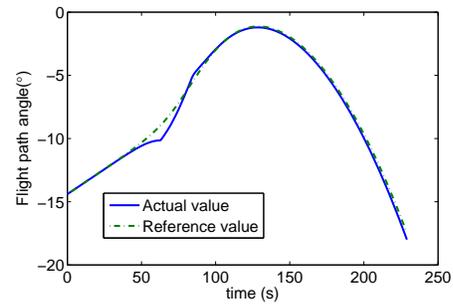}}
  \caption{Bank angle and flight path angel with state feedback guidance law (\ref{guidance law})\label{bankandsigma}}
\end{figure}

\begin{figure}[H]
  \centering
    \includegraphics[width=3in]{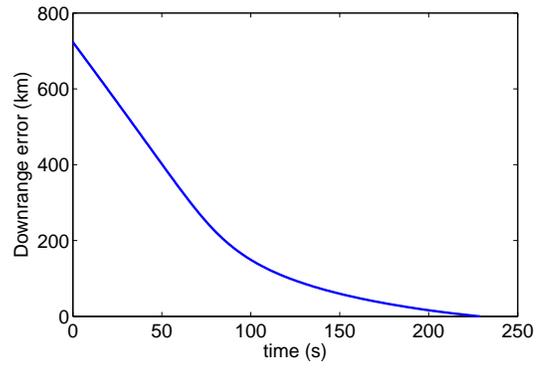}
  \caption{Downrange error with state feedback guidance law (\ref{guidance law})\label{downrange_error}}
\end{figure}

\begin{figure}[H]
  \centering
  \subfigure[Drag acceleration]{
    \includegraphics[width=2.5in]{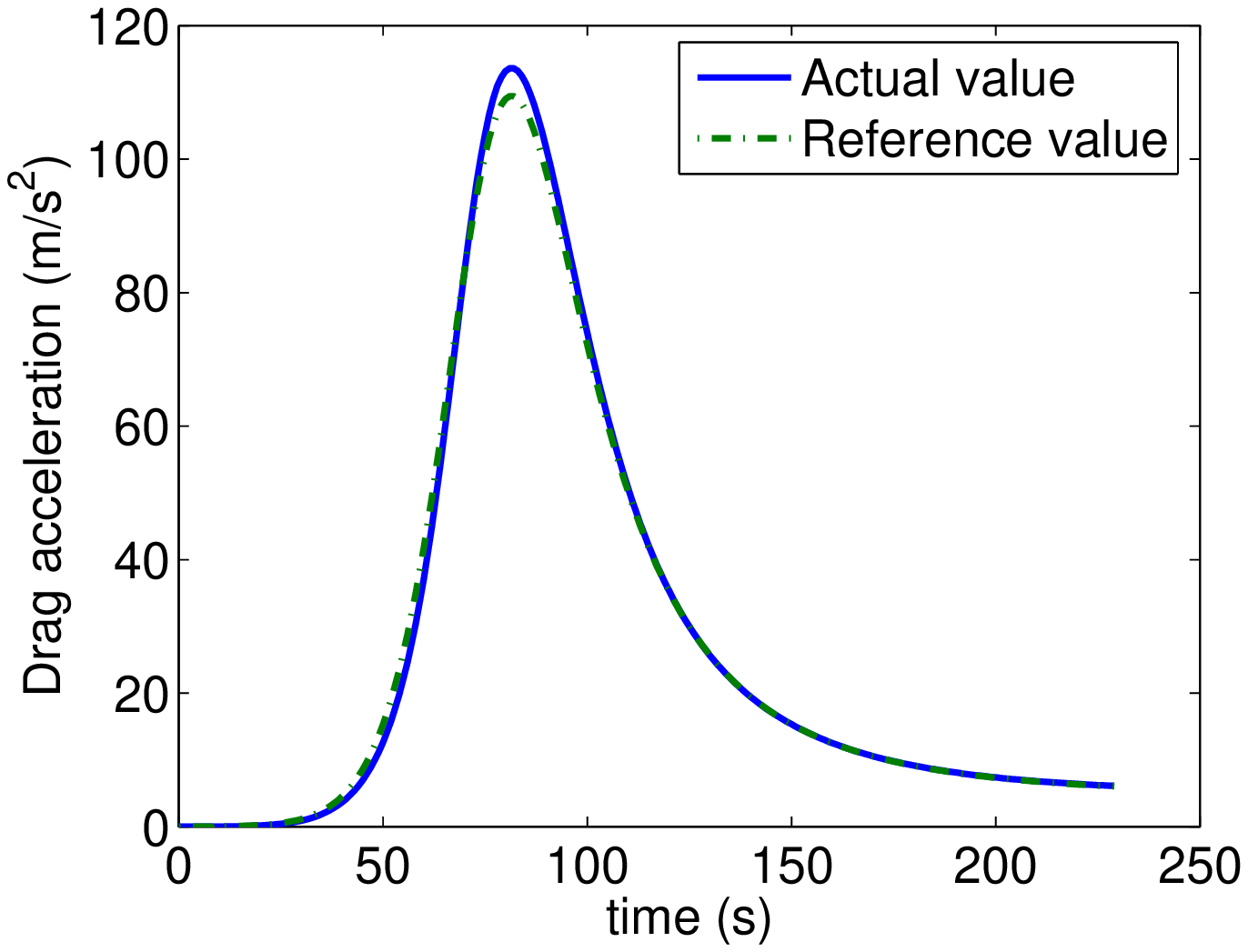}}
  \subfigure[Drag acceleration tracking error]{
    \includegraphics[width=2.5in]{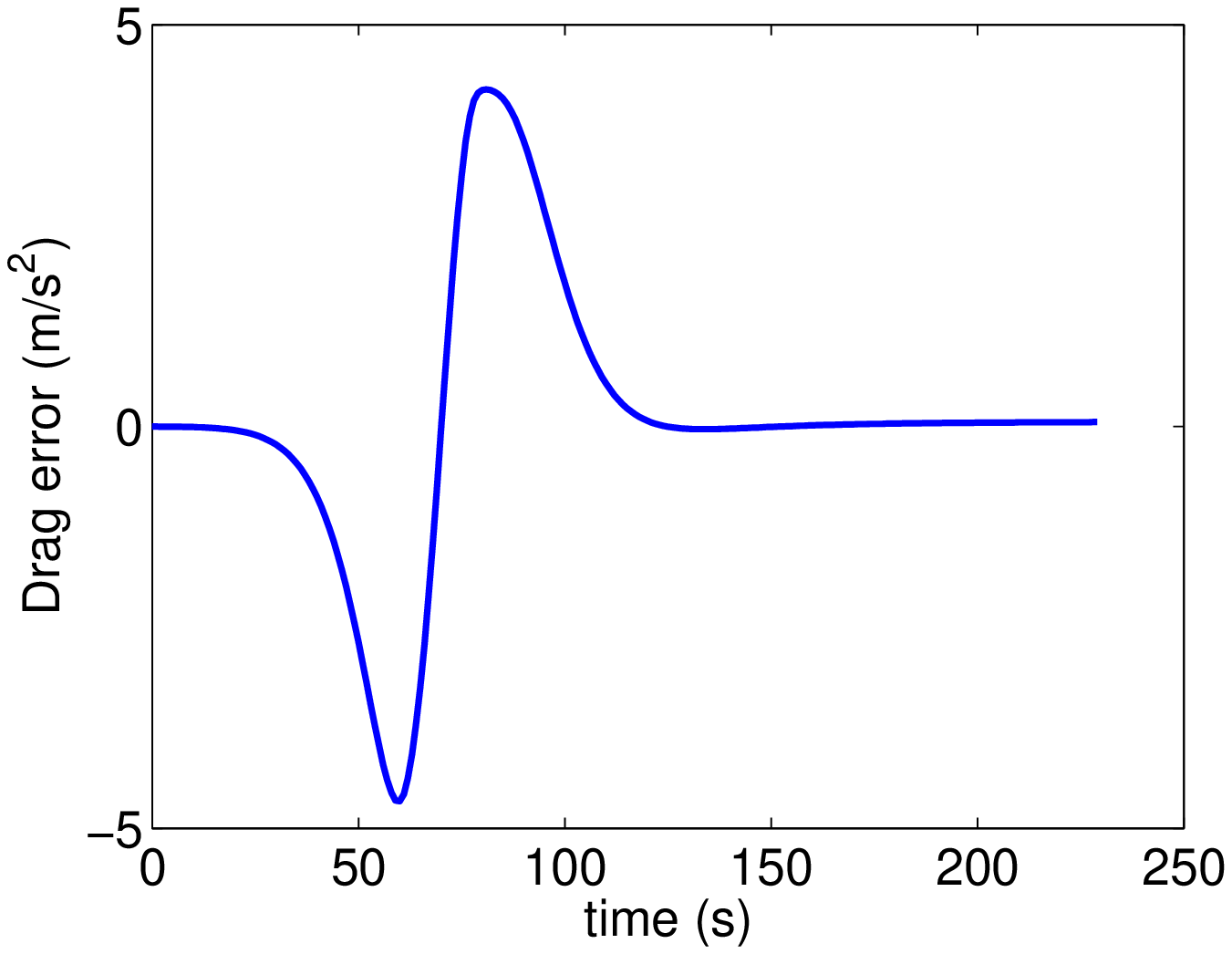}}
  \caption{Drag tracking with guidance law (\ref{guidance law with observer}) with observer (\ref{observer})\label{2drag}}
\end{figure}

\begin{figure}[H]
  \centering
  \subfigure[Altitude]{
    \includegraphics[width=2.5in]{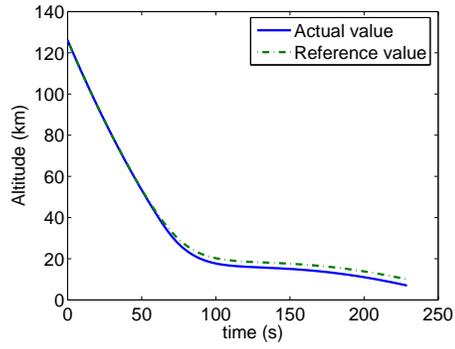}}
  \subfigure[Altitude tracking error]{
    \includegraphics[width=2.5in]{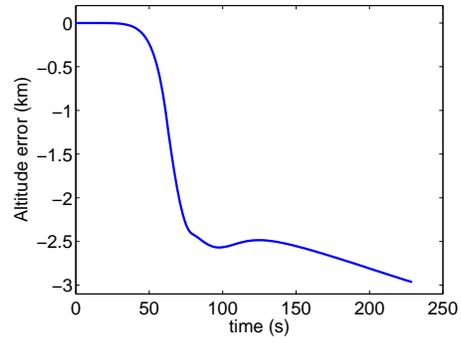}}
  \caption{Altitude tracking with guidance law (\ref{guidance law with observer}) with observer (\ref{observer})\label{2Altitude}}
\end{figure}

\begin{figure}[H]
  \centering
  \subfigure[Velocity]{
    \includegraphics[width=2.5in]{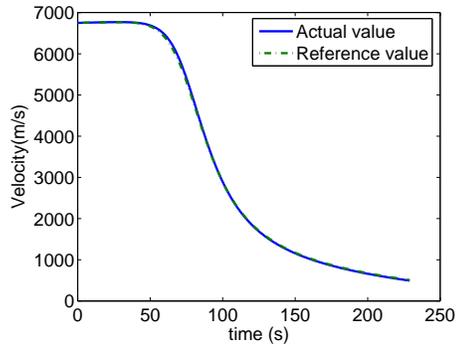}}
  \subfigure[Velocity tracking error]{
    \includegraphics[width=2.5in]{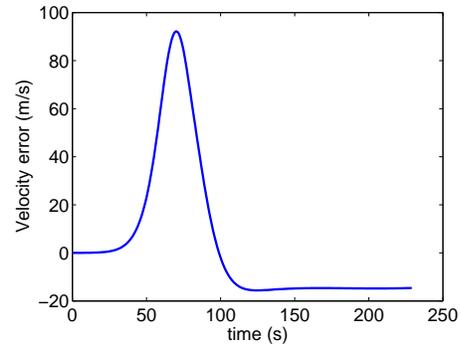}}
  \caption{Velocity tracking with guidance law (\ref{guidance law with observer}) with observer (\ref{observer})\label{2Velocity}}
\end{figure}

\begin{figure}[H]
  \centering
  \subfigure[$\sigma$]{
    \includegraphics[width=2.5in]{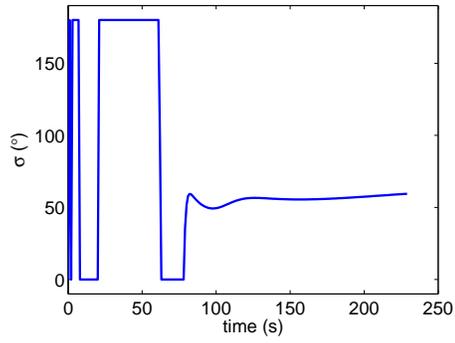}}
  \subfigure[$\gamma$]{
    \includegraphics[width=2.5in]{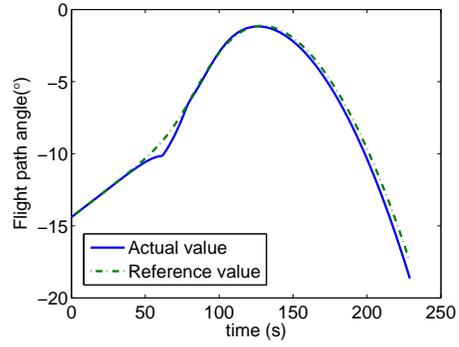}}
  \caption{Bank angle and flight path angel with guidance law (\ref{guidance law with observer}) with observer (\ref{observer})\label{2bankandsigma}}
\end{figure}

\begin{figure}[H]
  \centering
    \subfigure[Downrange error]{
    \includegraphics[width=2.5in]{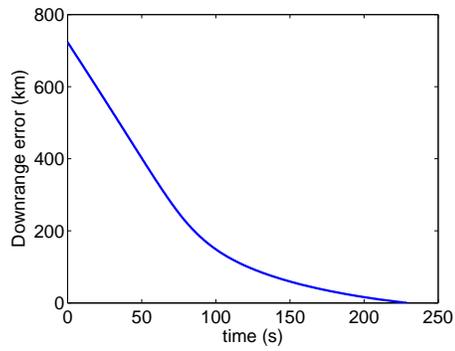}}
    \subfigure[Drag rate]{
    \includegraphics[width=2.5in]{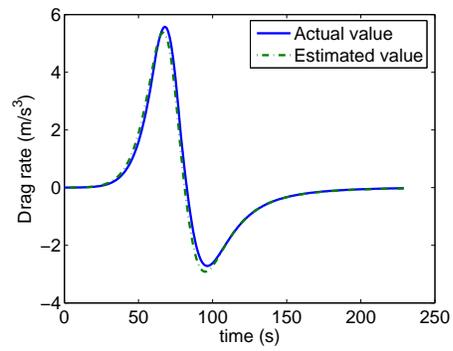}}
  \caption{Downrange error and drag rate with guidance law (\ref{guidance law with observer}) with observer (\ref{observer})\label{2downrange_error}}
\end{figure}

Fist, the performance of guidance law (\ref{guidance law}) is tested with taking $\varepsilon_{0}=5,a=1.982,b=3$, and the simulation results are shown in Figs. \ref{drag}-\ref{downrange_error}. It is can be seen that the reference profiles can be well tracked under the guidance law, and the downrange error is 0.00475km. Then, guidance law (\ref{guidance law with observer}) with observer (\ref{observer}) is used for $\varepsilon_{0}=5,a=1.982,b=3,l_{1}=2l_{2}=2,\varepsilon=0.481$, and the simulation results are shown in Figs. \ref{2drag}-\ref{2downrange_error}. Comparing with Figs. \ref{drag}-\ref{downrange_error}, we can see that the  performance of guidance law (\ref{guidance law}) can be recovered by using the high-gain observer with sufficiently small $\varepsilon$, and the downrange error is 0.0722km. Since the atmospheric density is very small at the beginning of entry and it leads to the fact that $g_{0}(D,t)=-\left(\frac{v}{h_{s}}+\frac{2g}{v}\right)\frac{LD\cos\gamma}{v}$ is small, thus, a large control magnitude is needed to make the drag track its reference value, which is the reason why bank angle reaches saturation level at initial time with both guidance laws.

\begin{table}[h]
\begin{center}
\caption{State Variables\label{State}}
\begin{tabular}{|c|c|}\hline
Initial State Variables & Final State Variables \\ \hline
\begin{tabular}{c|c}
Altitude, $h_0$ (km) & 126.1\\
Relative velocity, $V_0$ (km/s) & 6.75\\
Flight path angle,  $\gamma_{0}$ ($\circ$) & -14.4\\
Longitude ($\circ$) & 0\\
Latitude ($\circ$) & 0
\end{tabular}&
\begin{tabular}{c|c}
Altitude, $h_f$ (km) & 10 \\
Relative Velocity, $V_f$ (m/s) & 503 \\
Flight path angle,  $\gamma_{f}$ ($\circ$) & --- \\
Longitude ($\circ$) & 12.2 \\
Latitude ($\circ$) & 0
\end{tabular} \\ \hline
\end{tabular}
\end{center}
\end{table}

\begin{table}[h]
\begin{center}
\caption{Statistics of Dispersions Used in Monte Carlo Study\label{daba}}
\begin{tabular}{|c|c|c|}\hline
Parameters & Distribution & $[\Delta^{-},\Delta^{+}]$ \\ \hline
Mass deviation & uniform & [-5\%,5\%]\\
Atmospheric density deviation & uniform & [-20\%,20\%]\\
$C_{L}$ deviation & uniform & [-30\%,30\%]\\
$C_{D}$ deviation & uniform & [-30\%,30\%]\\ \hline
\end{tabular}
\end{center}
\end{table}

\begin{table}[h]
\begin{center}
\caption{Result of Monte Carlo Study\label{daba result}}
\begin{tabular}{|c|c|c|}\hline
 & Downrange Error (km) & Altitude Error (km) \\ \hline
Minimum & 0.0028 & -0.0013\\
Maximum & 24.3249 & 3.9263\\
Average & 2.3957 & 0.6369\\
Standard deviation & 6.7809 & 0.8584\\ \hline
\end{tabular}
\end{center}
\end{table}

\begin{figure}[H]
  \centering
    \subfigure[Downrange error]{
    \includegraphics[width=2.5in]{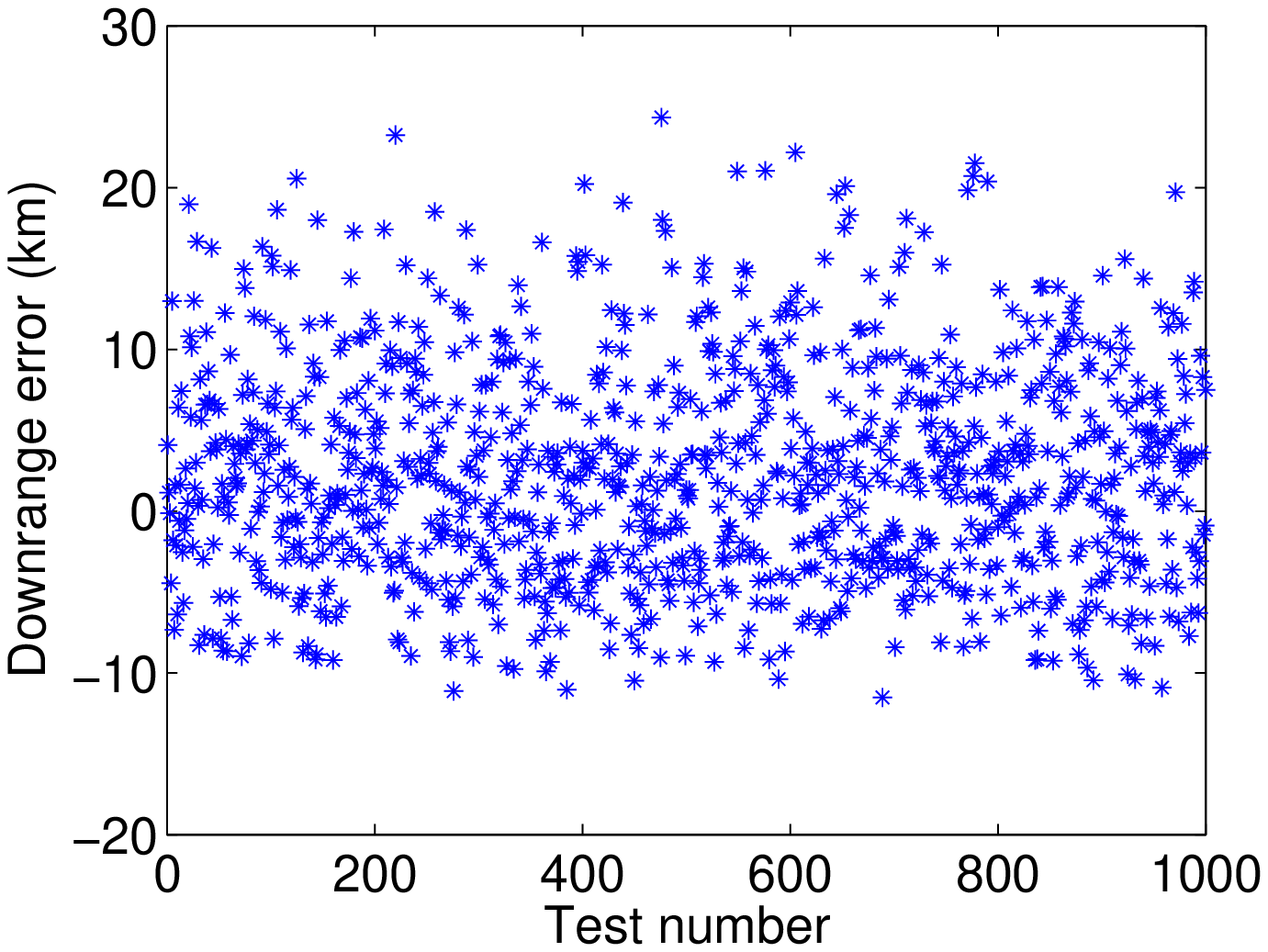}}
    \subfigure[Altitude error]{
    \includegraphics[width=2.5in]{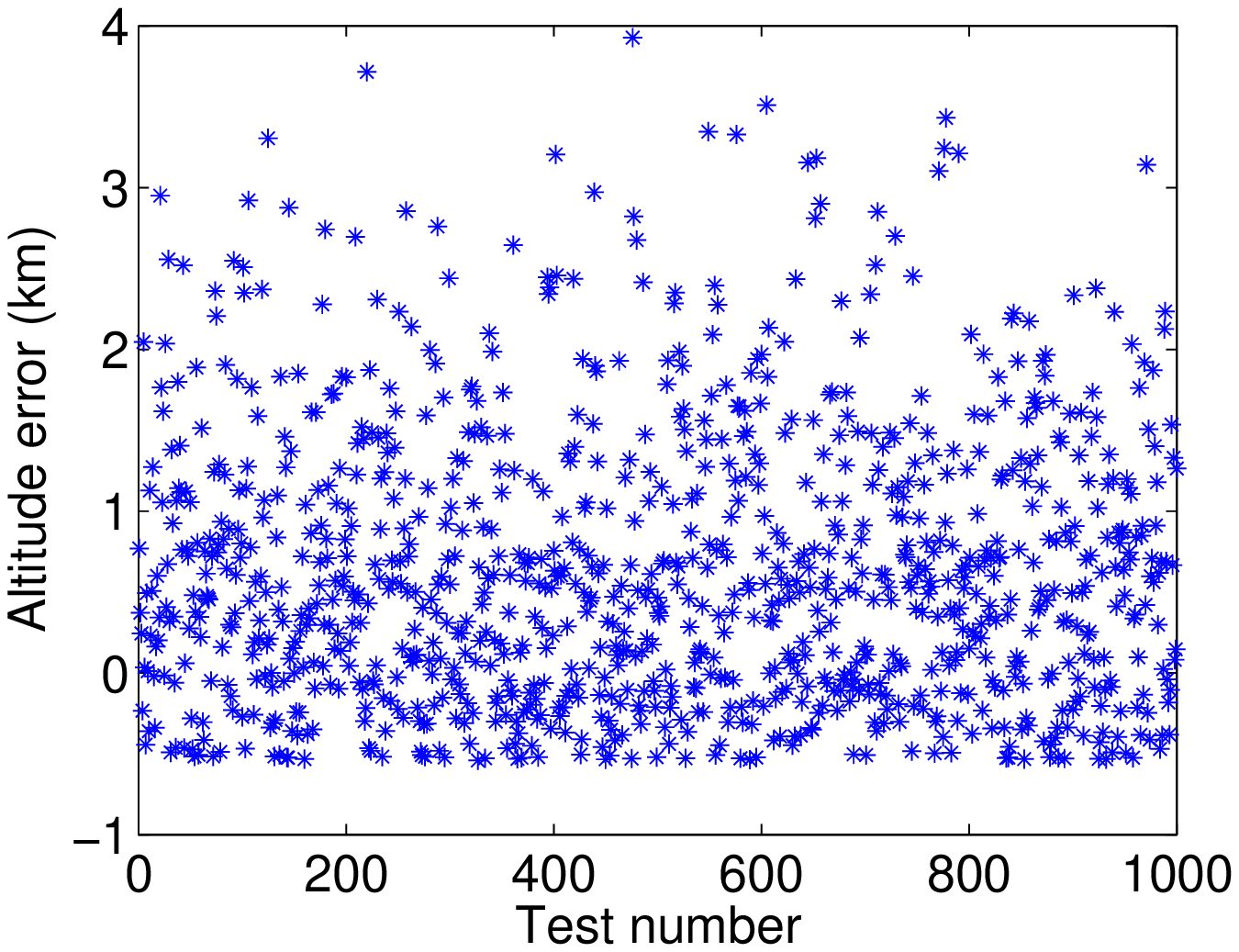}}
  \caption{Monte Carlo study of guidance law (\ref{guidance law with observer}) with observer (\ref{observer})\label{1000daba}}
\end{figure}

To test the robustness of the proposed guidance law (\ref{guidance law with observer}) with observer (\ref{observer}), a 1000-run Monte Carlo study using the parameter deviation in Table \ref{daba} is done. Take $\varepsilon_{0}=20,a=20,b=5,l_{1}=2l_{2}=2,\varepsilon=0.45$, and the result is shown in Fig. \ref{1000daba}. We can see that most of the downrange errors can be kept between -10km and 20km, while the altitude errors are kept between -0.6km and 4km. The result of this Monte Carlo study is summarized in Table \ref{daba result}.

\section{Conclusions}\label{conclusion}
A nonlinear drag-tracking guidance law was designed based on input-to-state stability (ISS) and a high-gain observer for entry vehicles. The proposed approach does not require prior information of drag rate, and it was proven that the drag-tracking error is ISS with respect to the uncertainties by using the guidance law with a sufficiently fast high-gain observer. The stability analysis as well as the simulation results show that the scheme can be effectively used in entry phase.

\section*{Acknowledgments}
The authors would like to thank Dr. Minwen Guo for her help in simulation study, and Dr. Xinghu Wang for his comments on this paper.

\begin{thebibliography}{0}
\bibitem{Guo_2013}
D. Y. Wang and M. W. Guo, Robust guidance law for drag tracking in mars atmospheric entry flight, Proceedings of the 33rd Chinese Control Conference, Nanjing, China, 2014, pp. 697-702.
\bibitem{Saraf_2004}
A. Saraf, A. Leavitt, D. T. Chen, et al. Design and evaluation of an acceleration guidance algorithm for entry, Journal of Spacecraft and Rockets, 2004, Vol. 41, No. 6, 2004, pp. 986-996.
\bibitem{Shuttle}
J. C. Harpold and C. A. Graves, Shuttle entry guidance, Journal of the Astronautical Sciences, Vol. 27, No. 3, 1979, pp. 239-268.
\bibitem{Mease_2002}
K. D. Mease, D. T. Chen, P. Teufel, et al. Reduced-order entry trajectory planning for acceleration guidance, Journal of Guidance, Control and Dynamics, 2002, Vol. 25, No. 2, 2002, pp. 257-266.
\bibitem{Leavitt_2007}
J. A. Leavitt, K. D. Mease, Feasible trajectory generation for atmospheric entry guidance, Journal of Guidance, Control and Dynamics, Vol. 30, No. 2, 2007 pp. 472-481.
\bibitem{Wang_2013}
S. H. Wang, P. L. Fei, X. X. Liu and B. Zhang, A new evolved acceleration reentry guidance for reusable launch vehicles, Applied Mechanics and Materials, Vols. 380-384, 2013, pp. 576-580.
\bibitem{Talole_2007}
S. E. Talole, J. Benito and K. D. Mease, Sliding mode observer for drag tracking in entry guidance, Proceedings of the AIAA Guidance, Navigation and Control Conference, Hilton Head, South Carolina, 2007, pp. 5122-5137.
\bibitem{Lu_1994}
P. Lu, Nonlinear Predictive controllers for continuous systems, Journal of Guidance, Control and Dynamics, Vol. 17, No. 3, 1994, pp. 553-560.
\bibitem{Lu_1997}
P. Lu, Entry guidance and trajectory control for reusable launch vehicle, Journal of Guidance, Control and Dynamics, Vol. 20, No. 1, 1997, pp. 143-149.
\bibitem{Benito_2008}
J. Benito and K. D. Mease, Nonlinear predictive controller for drag tracking in entry guidance, Proceedings of the 2008 AIAA/AAS Astrodynamics Specialist Conference and Exhibit, Honolulu, Hawaii, AIAA-2008-7350.
\bibitem{Guo}
M. W. Guo and D. Y. Wang, Guidance laws for low-lifting skip reentry subject to control saturation based on nonlinear predictive control, Aerospace Science and Technology, doi:10.1016/j.ast.2014.05.004.
\bibitem{Spong}
M. W. Spong, On the robust control of robot manipulators, IEEE Transactions on Automatic Control, Vol. 37, 1992, pp. 1782-1786.
\bibitem{Xia_2013}
R. F. Chen, Y. Q. Xia, Drag-based entry guidance for mars pinpoint landing, Proceedings of the 32nd Chinese Control Conference, Xi'an, China, 2013, pp. 5473-5478.
\bibitem{Tian_2011}
B. L. Tian and Q. Zong, Optimal guidance for reentry vehicle based on indirect Legendre pseudospectral method, Acta Astronautica, Vol. 68, No. 7-8, 2011, pp. 1176-1184.
\bibitem{Khalil_1992}
F. Esfandiari and H. K. Khalil, Output feedback stabilization of fully linearizable systems, International Journal of Control, Vol. 56, No. 5, 1992, pp. 1007-1037.
\bibitem{Khalil_1999}
A. N. Atassi and H. K. Khalil, A separation principle for the stabilization of a class of nonlinear systems, IEEE Transactions on Automatic Control, Vol. 44, No. 9, 1999, pp. 1672-1687.
\bibitem{book_of_K.Khalil}
H. K. Khalil, Nonlinear systems, 3rd ed., Prentice-Hall, Upper Saddle River, NJ, 2002, Chap. 4.
\bibitem{Mars}
H. J. Shen, H. Seywald and R. W. Powell, Desensitizing the pin-point landing trajectory on Mars, Proceedings of the AIAA/AAS Astrodynamics Specialist Conference and Exhibit, Honolulu, Hawaii, 2008.
\end{thebibliography}
\end{document}